\documentstyle[twocolumn,aps,epsfig]{revtex}

\begin{document}
\title{Strange Resonance Production: \\
Probing  Chemical and Thermal Freeze-out 
in Relativistic Heavy Ion Collisions}

\author{Marcus Bleicher and J\"org Aichelin}

\address{SUBATECH,
Laboratoire de Physique Subatomique et des
Technologies Associ\'ees \\
University of Nantes - IN2P3/CNRS - Ecole des Mines de Nantes \\
4 rue Alfred Kastler, F-44072 Nantes, Cedex 03, France}

\maketitle

\noindent
\begin{abstract}
The production and the observability of
$\Lambda(1520)$, $K^0(892)$ $\Phi$ and $\Delta(1232)$ 
hadron resonances in central Pb+Pb 
collisions at 160 AGeV is addressed. The rescattering probabilities
of the resonance decay products in the evolution are studied.
Strong changes in the reconstructable particle yields and 
spectra between chemical and thermal freeze-out are estimated. 
Abundances and spectra of reconstructable resonances are predicted.
\vspace{1cm}
\end{abstract}

Strange particle yields and spectra are key  probes
to study excited nuclear matter and to detect 
the transition of (confined) hadronic matter to 
quark-gluon-matter (QGP)\cite{raf8286,soff99plb,senger99,JPG,stock99,and98a,sikler99,cgreiner00}.
The relative enhancement of strange and  multi-strange
hadrons, as well as hadron ratios in central
heavy ion collisions with respect to peripheral or proton
induced interactions have been suggested as a signature
for the transient existence of a QGP-phase \cite{raf8286}.

Unfortunately, the emerging final state particles remember relatively
little about their primordial source, since they had been subject
to many rescatterings in the hadronic gas stage\cite{Soff:2000ae,PBM99,CERN}.
This has given rise to the interpretation of hadron production in terms
of thermal/statistical models. Two different kinds of freeze-outs
are assumed in these approaches: 
\begin{enumerate}
\item
a chemical freeze-out, were the 
inelastic flavor changing collisions processes cease,
roughly at an energy per particle of 1~GeV\cite{Cleymans:1998fq}, 
\item 
followed
by a later kinetic/thermal freeze-out were also elastic processes
have come to an end and the system decouples.
\end{enumerate}
Chemical and thermal freeze-out happen sequentially at 
different temperatures ($T_{\rm ch} \approx 160-170$~MeV, 
$T_{\rm th} \approx 120$~MeV) and thus at different times.

To investigate the sequential freeze-out in heavy ion reactions at SPS 
the Ultra-relativistic Quantum Molecular Dynamics model (UrQMD 1.2)
is applied \cite{urqmd1,urqmd2}.
UrQMD is a microscopic transport approach based on the 
covariant propagation of
constituent quarks and diquarks accompanied by mesonic and baryonic
degrees of freedom.  It simulates multiple interactions of ingoing
and newly produced particles, the excitation and fragmentation of
color strings and the formation and decay of hadronic resonances.  At
present energies, the treatment of sub-hadronic degrees of freedom is of
major importance.  In the UrQMD model, these degrees of freedom enter
via the introduction of a formation time for hadrons produced in the
fragmentation of strings \cite{andersson87a,andersson87b,sjoestrand94a}.  
The leading hadrons
of the fragmenting strings contain the valence-quarks of the original
excited hadron. In UrQMD they are allowed to interact even during
their formation time, with a reduced cross section defined by the
additive quark model, thus accounting for the original valence quarks
contained in that hadron \cite{urqmd1,urqmd2}. Those leading hadrons
therefore represent a simplified picture of the leading (di)quarks of
the fragmenting string.  Newly produced (di)quarks do, in the present
model, not interact until they have coalesced into hadrons --
however, they contribute to the energy density of the system.  
For further details about the UrQMD model,
the reader is referred to Ref. \cite{urqmd1,urqmd2}.

\begin{figure}[h]
\vskip 0mm
\vspace{0cm}
\centerline{\psfig{figure=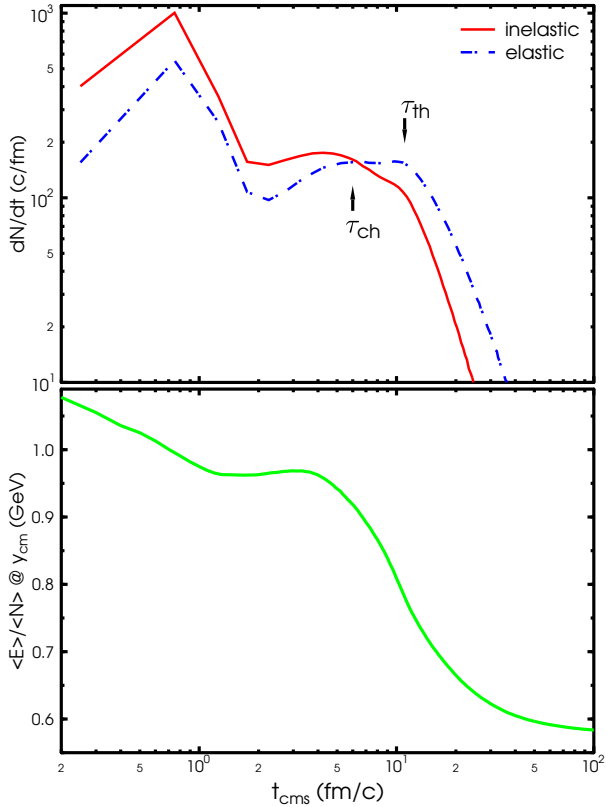,width=3.5in}}
\vskip 2mm
\caption{Top:Inelastic and (pseudo-)elastic collision rates in Pb+Pb at 160AGeV. 
$\tau_{\rm ch}$ and $\tau_{\rm th}$ denote the chemical and thermal/kinetic freeze-out as
given by the microscopic reaction dynamics of UrQMD. Bottom: Average energy per particle 
at midrapidity ($|y-y_{\rm cm}|\le 0.1$) as a function of time.
\label{chf}}
\end{figure}

Let us start by asking whether a microscopic non-equilibrium model can support
the ideas of sequential chemical and thermal break-up of the hot and dense matter?
To analyze the different stages of a heavy ion collision, Fig. \ref{chf}
shows the time evolution of the elastic and inelastic collision rates in Pb+Pb
at 160AGeV. The inelastic collision rate (full line) is defined as the number of
collisions with flavor changing processes (e.g. $\pi\pi\rightarrow K\overline K$). 
The elastic collision rate (dashed dotted line) consists of two components, 
true elastic processes (e.g. $\pi\pi\rightarrow \pi\pi$) and pseudo-elastic 
processes (e.g. $\pi\pi\rightarrow \rho \rightarrow \pi\pi$).
While elastic collision do not change flavor, pseudo-elastic collisions are different.
Here, the ingoing hadrons are destroyed and a resonance is formed. If this resonance
decays later into the same flavors as its parent hadrons, this scattering 
is pseudo-elastic. 

Indeed the main features revealed by the present microscopic study do not contradict the 
idea of a chemical and thermal break-up of the source as shown in Fig. \ref{chf} (top)
However, the detailed freeze-out dynamics is much richer and by far more complicated as
expected in simplified models:
\begin{enumerate}
\item
In the early non-equilibrium stage of 
the AA collision ($t<2$~fm/c)  the collision rates are huge and vary strongly with time.
\item
The intermediate stage (2~fm/c$<t<6$~fm/c) is dominated by inelastic, flavor and chemistry 
changing processes until chemical freeze-out. 
\item
This regime is followed by a phase of dominance 
of elastic and pseudo-elastic collisions (6~fm/c$<t<11$~fm/c). Here only the momenta
of the hadrons change, but the chemistry of the system is mainly unaltered, 
leading to the thermal freeze-out of the system. 
\end{enumerate}
Finally the system breaks-up ($t>11$~fm/c)
and the scattering rates drop exponentially.

Fig. \ref{chf} (bottom) depicts the average energy\footnote{To compare to thermal model estimates, 
the energies are calculated from interacted hadrons
with the assumption $p^2_z=\left(p^2_x+p^2_y\right)/2$. This assures independence of the 
longitudinal motion in the system and the chosen rapidity cut.}
per particle at midrapidity ($|y-y_{\rm cm}|\le 0.1$). 
One clearly observes a correlation between chemical break-up in terms of inelastic scattering
rates and the rapid decrease in energy per particle. Thus, the suggested phenomenological 
chemical freeze-out condition of $\approx 1$~GeV/particle is also found in the present 
microscopic model analysis.

To verify this scenario, we exploit 
the spectra and abundances of $\Lambda(1520)$, $K^0(892)$ and other resonances 
which unravel the break-up dynamics of the source between 
chemical and thermal freeze-out.
In the statistical model interpretation of heavy ion reactions the 
resonances are produced at chemical freeze-out. If chemical and thermal
freeze-out are not separated - e.g. due to an explosive break-up of the
source - all initially produced resonances are reconstructable by an 
invariant mass analysis in the final state. However, if there is a separation between the
different freeze-outs, a part of the resonance daughters rescatter, making this 
resonance unobservable in the final state. Thus, the relative suppression of
resonances in the final state compared to those expected from thermal
estimates provides a chronometer for the time period between the different
reaction stages.
Even more interesting, inelastic scatterings 
of the resonance daughters (e.g. $\overline K p \rightarrow \Lambda$) might change the
chemical composition of the system {\it after} 'chemical' freeze-out by as much as 10\% for
all hyperon species. 
While this is not in line with the thermal/statistical model interpretation, it 
supports the more complex freeze-out dynamics encountered in the present model.

To answer these questions we address the experimentally accessible
hadron resonances:
At this time  $\Phi$ and $\Lambda(1520)$ have
been observed in heavy ion reactions at SPS energies
 \cite{NA49Res,NA49Res2} following a suggestion
that such a measurement was possible \cite{Bec98}.
SPS \cite{NA49Res2} and RHIC experiments \cite{STARkstar}
report measurement of the $\overline{K^0}(892)$ signal, and
RHIC has already measured both the $K^0$ and the
$\overline{K^0}$.  In the SPS case,
the $\Lambda(1520)$ abundance yield is about 2.5 times smaller
than expectations based on the yield
extrapolated from nucleon-nucleon reactions. This is of highest interest
in view of the $\Lambda$ enhancement by factor 2.5 of
in the same reaction compared to elementary collisions.

As an  explanation for
this effective suppression by a factor 5, we show that the
decay products ($\pi,\Lambda$, etc.) produced at rather high
chemical freeze-out temperatures and densities  have rescattered. 
Thus, their momenta do not allow  to reconstruct these states 
in an invariant mass analysis.
However, even the question of the existence of
such resonance states in the hot and dense environment is still 
not unambiguously answered. Since hyperon resonances are
expected to dissolve at high energy densities (see .e.g. \cite{Lutz:2001dq}) 
it is of utmost importance to study the cross section 
of hyperon resonances as a thermometer of the collision. 

The present exploration considers the resonances,
$\Delta(1232)$, $\Lambda^*(1520)$, K$^0(892)$ 
and $\Phi$. 
The properties of these hadrons  are depicted in 
Table \ref{a}. 
\begin{table}[h!]
\begin{tabular}{lrrr}
Particle     & Mass (MeV) & Width (MeV) & $\tau$ (fm/c)\\\hline
$\Delta(1232)$     & 1232 & 120  & 1.6   \\
$\Lambda^*(1520)$  & 1520 & 16   &  12   \\
K$^0(892)$         & 893  & 50   &  3.9   \\
$\Phi(1020)$       & 1019 & 4.43 & 44.5    \\
\end{tabular}
\caption{Properties of investigated resonances.}
\label{a}
\end{table} 

\begin{figure}[h]
\vskip 0mm
\vspace{0cm}
\centerline{\psfig{figure=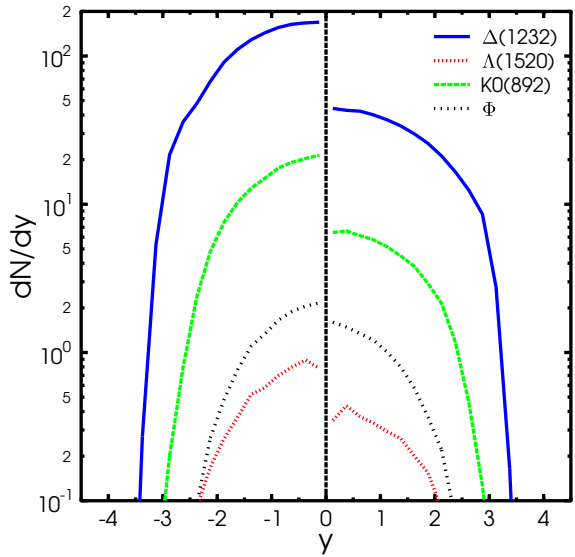,width=3.5in}}
\vskip 2mm
\caption{Rapidity densities for $\Delta(1232)$, $\Lambda^*(1520)$, K$^0(892)$ 
and $\Phi$ in Pb($160\,A$GeV)Pb, b$<3.4$~fm collisions.
Left: All resonances as they decay. Right: Reconstructable resonances.
\label{dndy}}
\end{figure}

Fig. \ref{dndy} shows the rapidity densities for 
$\Delta(1232)$, $\Lambda^*(1520)$, K$^0(892)$ 
and $\Phi$ in Pb($160\,A$GeV)Pb, b$<3.4$~fm collisions.
Fig. \ref{dndy} (left), shows the total amount of decaying resonances. 
Here, subsequent collisions of the decay products have 
not been taken into account - i.e. whenever a resonance decays during 
the systems evolution it is counted.
However, the additional interaction of the daughter hadrons 
disturbs the signal of the resonance in the invariant mass spectra.
This lowers the observable yield of resonances drastically as 
compared to the primordial yields at chemical freeze-out.
Fig. \ref{dndy}(right) addresses this in the rapidity distribution of those
resonances whose decay products do not suffer subsequent collisions - 
these resonances are in principle reconstructable from their decay products.
Note that reconstructable in this context still assumes reconstruction of
all decay channels, including many body decays.

At SPS energies, the rapidity distributions d$N$/d$y$ may by described by a Gaussian
curve 
\begin{equation}
\frac{{\rm d}N}{{\rm d}y}(y) 
= A \times {\rm exp}\left(-\frac{y^2}{2\sigma^2}\right)\quad.
\end{equation}
with parameters given in Table \ref{tab1}.
However, in this approximation the details of
the rapidity distributions are lost. Especially the
dip in the rapidity distributions of the $\Lambda(1520)$ is 
not accounted for. 

The rescattering probability of the resonance decay products depends
on the cross section of the decay product with the surrounding matter, on the lifetime
of the surrounding hot and dense matter, on the lifetime of the resonance
and on the specific properties of the daughter hadrons in the resonance decay channels.
This leads to different 'observabilities' of the different resonances:
Rescattering influences the observable $\Phi$ and $\Lambda^*$ yields 
only by a factor of two, due to the long life time of those particles.
In contrast strong effects are observed in the K$^*$ and $\Delta$ 
yields, which are suppressed by more than a  factor three.

One can use the estimates done by \cite{Torrieri:2001ue} in a statistical 
model, and try to relate the result of the present microscopic 
transport calculation to thermal freeze-out parameters.
The ratios of the $4\pi$ numbers of 
reconstructable is $\Lambda(1520)/\Lambda=0.024$ and K$^0(892)/$K$^+=0.25$. 
In terms of the analysis by \cite{Torrieri:2001ue}, 
the microscopic source has a lifetime below 1~fm/c and a freeze-out 
temperature  below 100~MeV. Thus, the values obtained from UrQMD seem to 
favor a scenario of a sudden 
break-up of the initial hadron source, in contrast to the time evolution 
of the chemical and thermal decoupling as shown in Fig. \ref{chf}.
However, note that these numbers are based on the above mentioned 
thermal scenario. In fact, fitting hadron ratios of UrQMD calculations for central 
Pb+Pb interactions at 160~AGeV with a Grand Canonical ensemble yields
a chemical freeze-out temperature of $150-160$~MeV \cite{Bass:1998xw}.

\begin{figure}[h]
\vspace{-0.5cm}
\centerline{\psfig{figure=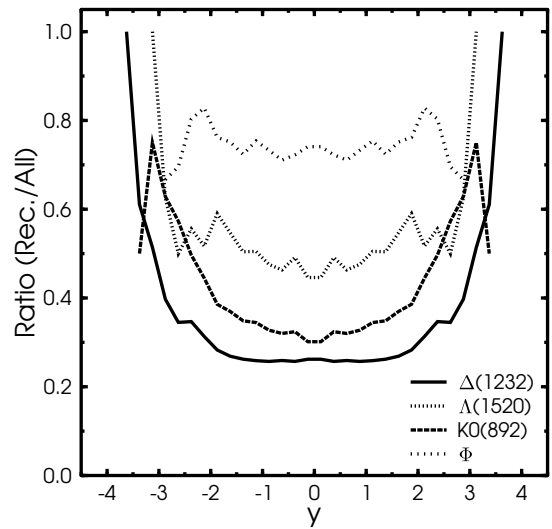,width=3.5in}}
\caption{Rapidity dependent ratio $R$ of reconstructable resonances over all resonances 
of a given type as they decay. For $\Delta(1232)$, $\Lambda^*(1520)$, 
K$^0(892)$ and $\Phi$ in Pb($160$~AGeV)+Pb, b$<3.4$~fm collisions.
\label{ratio}}
\end{figure}

In fact, the rescattering strength depends on the phase space region studied.
Fig. \ref{ratio} addresses the longitudinal momentum distribution of
the rescattering strength. The probability $R$ to observe the resonance $H$ 
is given by
\begin{equation}
R=\frac{H \rightarrow h_1 h_2 ({\rm reconstructable})}
{H \rightarrow h_1 h_2 ({\rm all})}\quad.
\end{equation}
The 'observability' decreases strongly towards central rapidities.
This is due to the higher hadron density at central rapidities which 
increases the absorption probability of daughter hadrons drastically.
Unfortunately, in the case of the $\Phi$ meson this decrease of 
the observable yield increases the discrepancies between data and
microscopic model predictions (e.g.\cite{soff99plb}).

\begin{figure}[h]
\vskip 0mm
\vspace{0cm}
\centerline{\psfig{figure=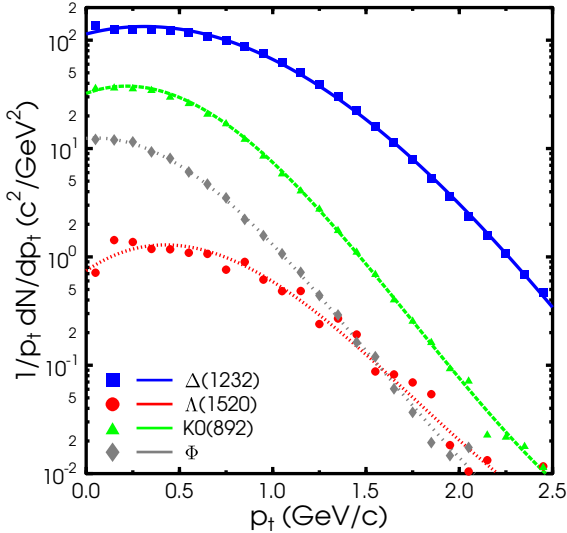,width=3.5in}}
\vskip 2mm
\caption{Transverse momentum spectra at $|y-y_{\rm cm}|<1$ of reconstructable 
resonances for $\Delta(1232)$, $\Lambda^*(1520)$, 
K$^0(892)$ and $\Phi$ in Pb($160\,A$GeV)Pb, b$<3.4$~fm collisions.
The lines are to guide the eye.
\label{dndpt}}
\end{figure}

The absorption probability of daughter hadrons is not only rapidity
dependent but also transverse momentum dependent.
The decay products are rescattered preferentially at low transverse momenta. 
Fig. \ref{dndpt} depicts the invariant transverse momentum spectra of
reconstructable $\Delta(1232)$, $\Lambda^*(1520)$, K$^0(892)$ 
and $\Phi$ in Pb($160\,A$GeV)Pb, b$<3.4$~fm collisions.
\begin{table}
\begin{tabular}{lrr}
Particle (All)           & A    & $\sigma$\\\hline
$\Delta$      & 181  & 1.49\\
$\Lambda^*$   & 0.89 & 1.20\\
K$^*$         & 22.0 & 1.23\\
$\Phi$        & 2.22 & 1.09\\\hline
Particle (Rec.)          & A    & $\sigma$\\\hline
$\Delta$      & 46.3 & 1.60\\
$\Lambda^*$   & 0.42 & 1.27\\
K$^*$         & 6.95 & 1.34\\
$\Phi$        & 1.62 & 1.10\\
\end{tabular}
\caption{Gaussian fit parameters for the rapidity densities of all and reconstructable 
resonances.}
\label{tab1}
\end{table}

\begin{figure}[h]
\vskip 0mm
\vspace{0cm}
\centerline{\psfig{figure=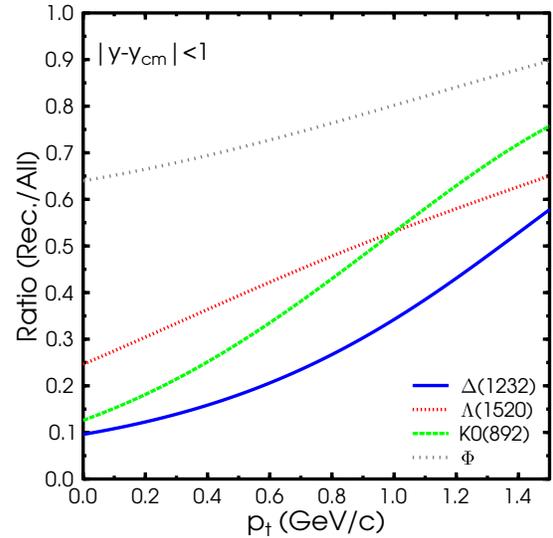,width=3.5in}}
\vskip 2mm
\caption{Ratio $R$ of reconstructable resonances over all resonances 
of a given type as a function of transverse momentum. 
For $\Delta(1232)$, $\Lambda^*(1520)$, 
K$^0(892)$ and $\Phi$ in Pb($160\,A$GeV)Pb, b$<3.4$~fm collisions.
\label{ratiopt}}
\end{figure}

Fig. \ref{ratiopt} directly addresses the  $p_t$ dependence of the
observability of resonances. 
The present model study supports a strong $p_t$ dependence of the
rescattering probability. This effects leads to a larger apparent 
temperature (larger inverse slope parameter) for resonances
reconstructed from strongly interacting particle.
A similar behavior has been found for the $\Phi$ meson in an 
independent study by \cite{johnson00}.
This effective heating of the $\Phi$ spectrum might explain 
the different inverse slope parameter measured by  
NA49 ($\phi\rightarrow K^+K^-$) as compared to the NA50 ($\phi\rightarrow \mu^+\mu^-$) 
collaboration \cite{soff99plb,johnson00,sikler99,willis99}.

In conclusion, central Pb+Pb interactions at 160 AGeV are studied 
within a microscopic non-equilibrium approach. The calculated scattering 
rates exhibit signs of a chemical and a subsequent
thermal freeze-out. The time difference between both freeze-outs 
is explored with hadronic resonances.
The rapidity and transverse momentum distributions
of strange and non-strange resonances are predicted. 
The observability of unstable (strange) particles 
($\Lambda(1520)$, $\phi$, etc.) in the invariant mass analysis 
of strongly interacting decay products is distorted due to
rescattering of decay products from chemical to thermal freeze-out.
Approximately $25\%$ of $\phi$'s 
and $50\%$ of $\Lambda^*$'s are not directly detectable by reconstruction 
of the invariant mass spectrum. The rescattering strength 
is strongly rapidity dependent.
Rescattering of the decay products alters the 
transverse momentum spectra of reconstructed resonances. 
This leads to higher apparent temperatures for resonances.
Inelastic collisions of anti-Kaons from decayed strange resonance
alter the chemical composition of strange baryons by up to 10\%.
{\tiny
\begin{table}[h!]
\begin{tabular}{ll}
 events  &1400\\
 $\Delta$/event&654.46\\
 $\Lambda^*$/event&2.60\\
 $K^0(892)$/event&66.2\\
 $\Phi$/event&5.917\\
\end{tabular}
\caption{$4\pi$ yields of all decaying resonances in Pb(160AGeV)+Pb,
b$<3.4$~fm.}
\end{table}
\begin{table}[h!]
\begin{tabular}{lrrrr}
y & dn/dy($\Delta$) & dn/dy($\Lambda(1520)$) & dn/dy(K$^0(892)$) & dn/dy($\Phi$)\\\hline
0.125 & 169.12 & 0.78 & 21.36 & 2.16\\
0.375 & 167.40 & 0.89 & 20.36 & 2.05\\
0.625 & 163.89 & 0.79 & 19.20 & 1.89\\
0.875 & 155.86 & 0.7  & 17.56 & 1.65\\
1.125 & 143.33 & 0.58 & 14.99 & 1.33\\
1.375 & 128.78 & 0.52 & 12.83 & 1.10\\
1.625 & 110.98 & 0.36 & 10.31 & 0.76\\
1.875 & 90.976 & 0.26 & 7.554 & 0.48\\
2.125 & 67.511 & 0.16 & 4.815 & 0.26\\
2.375 & 47.786 & 0.09 & 2.378 & 0.09\\
2.625 & 36.04  & 0.03 & 0.802 & 0.03\\
2.875 & 21.583 & 0.01 & 0.2   & 0.00\\
3.125 & 5.3613 & 0.00 & 0.028 & 0   \\
3.375 & 0.2719 & 0    & 0.00  & 0   \\
3.625 & 0.0017 & 0    & 0     & 0\\
3.875 & 0      & 0    & 0     & 0\\
\end{tabular}
\caption{Rapidity density of all decaying resonances in Pb(160AGeV)+Pb, b$<3.4$~fm.}
\end{table}
\newpage
\begin{table}[h!]
\begin{tabular}{ll}
 events  &1400\\
 $\Delta$/event&178.9\\
 $\Lambda^*$/event&1.28\\
 $K^0(892)$/event&22.61\\
 $\Phi$/event&4.35\\
\end{tabular}
\caption{$4\pi$ yields of all reconstructable resonances in Pb(160AGeV)+Pb,
b$<3.4$~fm.}
\end{table}
\begin{table}[h!]
\begin{tabular}{lrrrr}
y & dn/dy($\Delta$) & dn/dy($\Lambda(1520)$) & dn/dy(K$^0(892)$) & dn/dy($\Phi$)\\\hline
0.125 & 44.30 & 0.348 & 6.44 & 1.601\\
0.375 & 43.03 & 0.437 & 6.59 & 1.484\\
0.625 & 42.46 & 0.37  & 6.14 & 1.342\\
0.875 & 40.05 & 0.332 & 5.75 & 1.21\\
1.125 & 37.08 & 0.292 & 5.16 & 1.008\\
1.375 & 33.74 & 0.262 & 4.47 & 0.797\\
1.625 & 29.82 & 0.200 & 3.81 & 0.579\\
1.875 & 25.72 & 0.154 & 2.91 & 0.361\\
2.125 & 21.14 & 0.087 & 2.14 & 0.22\\
2.375 & 16.57 & 0.050 & 1.18 & 0.0752\\
2.625 & 12.43 & 0.018 & 0.46 & 0.022\\
2.875 & 8.570 & 0.007 & 0.12 & 0.002\\
3.125 & 2.755 & 0.001 & 0.02 & 0\\
3.375 & 0.165 & 0     & 0.00 & 0\\
3.625 & 0.001 & 0     & 0    & 0\\
3.875 & 0     & 0     & 0    & 0\\
\end{tabular}
\caption{Rapidity density of reconstructable resonances in Pb(160AGeV)+Pb,
b$<3.4$~fm.}
\end{table}
}


\begin{thebibliography}{99}
\bibitem{raf8286}
Rafelski~J, M\"uller~B
Phys. Rev. Lett.  {\bf 48}, (1982) 1066; (E) {\bf 56} (1986) 2334;\\
Koch~P, M\"uller~B, Rafelski~J
{ Phys.~Rep.} {\bf 142}, (1986) 167;\\
Koch~P, M\"uller~B, St\"ocker~H, Greiner~W
{ Mod.~Phys.~Lett.}~{\bf A3}, (1988) 737

\bibitem{soff99plb}
Soff~S, Bass~S~A, Bleicher~M, Bravina~L, Gorenstein~M,
Zabrodin~E, St\"ocker~H, Greiner~W,
Phys. Lett.~{\bf B471}, (1999) 89 

\bibitem{senger99}
Senger~P, Str\"obele~H
J. Phys. G {\bf 25}, (1999) R59

\bibitem{JPG}
{\it Strangeness in Quark Matter} (Padua, Italy, 1998),
J. Phys. G {\bf 25}, (1999) 143;
{\it Strangeness in Quark Matter} (Santorini, Greece, 1997),
J. Phys. G {\bf 23}, (1997) 1785;
{\it Relativistic Aspects of
Nuclear Physics} (Rio, Brazil, 1995),
T. Kodama et al., eds., World Scientific.
{\it Strangeness in Hadronic Matter} (Budapest, Hungary, 1996),
{\it Budapest, Akademiai Kiado}.
{\it Strangeness in Hadronic Matter} (Tucson, AZ, 1995),
{\it AIP Conf.} {\bf 340};
{\it Strange Quark Matter in Physics and Astrophysics} (Aarhus, Denmark, 1991),
Nucl. Phys. B{\bf 24B}, (1991)

\bibitem{stock99}
Stock~R, Phys. Lett. {\bf B456}, (1999) 277

\bibitem{and98a}
Andersen~E et al. [WA97 collaboratio]),
Phys. Lett.~{\bf B433}, (1998) 209;\\
Lietava~R et al. [WA97 collaboration],
J. Phys. G~{\bf 25}, (1999) 181;\\
Caliandro~R et al. [WA97 collaboration],
J. Phys. G~{\bf 25}, (1999) 171;\\
Margetis~S et al. [NA49 collaboration],
J. Phys. G~{\bf 25}, (1999) 189


\bibitem{sikler99}
J.~Bachler {\it et al.}  [NA49 Collaboration],
Nucl.\ Phys.\ A {\bf 661} (1999) 45.\\
C.~Hohne  [NA49 Collaboration],
Nucl.\ Phys.\ A {\bf 661} (1999) 485.

\bibitem{cgreiner00}
C.~Greiner and S.~Leupold,
J.\ Phys.\ G {\bf 27} (2001) L95
[arXiv:nucl-th/0009036].

\bibitem{Soff:2000ae}
S.~Soff {\it et al.},
J.\ Phys.\ G {\bf 27} (2001) 449
[arXiv:nucl-th/0010103].

\bibitem{PBM99}
P. Braun-Munzinger, I. Heppe and J. Stachel, {Phys. Lett.} B
\textbf{465}, 15 (1999).

\bibitem{CERN}
U.~W.~Heinz and M.~Jacob,
arXiv:nucl-th/0002042.\\
See also: \\ http://www.cern.ch/CERN/Announ\-cements/2000/NewStateMatter/

\bibitem{Cleymans:1998fq}
J.~Cleymans and K.~Redlich,
Phys.\ Rev.\ Lett.\  {\bf 81} (1998) 5284
[arXiv:nucl-th/9808030].

\bibitem{urqmd1} 
M. Bleicher, E. Zabrodin, C. Spieles, S.A. Bass,
C. Ernst, S. Soff, L. Bravina, M. Belkacem, H. Weber, H. St\"ocker,
W. Greiner, J. Phys. G 25 (1999) 1859.

\bibitem{urqmd2}
S.A. Bass, M. Belkacem,
M. Bleicher, M. Brandstetter, L. Bravina, C. Ernst, L. Gerland,
M. Hofmann, S. Hofmann, J. Konopka, G. Mao, L. Neise, S. Soff,
C. Spieles, H. Weber, L.A. Winckelmann, H. St\"ocker, W. Greiner,
C. Hartnack, J. Aichelin, N. Amelin,  Progr. Nucl. Phys. 41 (1998)
225. 

\bibitem{andersson87a} B.~Andersson, G.~Gustavson, and
B.~Nilsson-Almquist,  Nucl. Phys. {\bf B281}, 289 (1987).

\bibitem{andersson87b} B.~Andersson {\em et~al.}, 
Comp. Phys. Comm. {\bf 43}, 387 (1987).

\bibitem{sjoestrand94a} T.~Sjoestrand, 
Comp. Phys. Comm. {\bf 82}, 74 (1994).

\bibitem{NA49Res}
Ch. Markert [NA49 collaboration], \\ http://www.rhic.bnl.gov/qm2001/ ;\\
Ch. Markert, PhD thesis, Univ. Frankfurt, \\ http://na49info.cern.ch/cgi-bin/wwwd-util/NA49/NOTE?257

\bibitem{NA49Res2}
V. Friese [NA49 collaboration], \\ http://www.rhic.bnl.gov/qm2001/

\bibitem{Bec98}
F. Becattini, M. Ga\'zdzicki, and J. Solfrank
\textit{Eur. Phys. J.} C \textbf{5}, 143 (1998).

\bibitem{STARkstar}
Z.~b.~Xu  [STAR Collaboration],
Nucl.\ Phys.\ A {\bf 698} (2002) 607
[arXiv:nucl-ex/0104001].\\
H. Caines [STAR collaboration], \\ http://www.rhic.bnl.gov/qm2001/

\bibitem{Lutz:2001dq}
M.~F.~Lutz and C.~L.~Korpa,
arXiv:nucl-th/0105067.

\bibitem{Torrieri:2001ue}
G.~Torrieri and J.~Rafelski,
Phys.\ Lett.\ B {\bf 509} (2001) 239
[arXiv:hep-ph/0103149].


\bibitem{Bass:1998xw}
S.~A.~Bass {\it et al.},
Phys.\ Rev.\ Lett.\  {\bf 81} (1998) 4092
[arXiv:nucl-th/9711032].

\bibitem{johnson00}
S.~C.~Johnson, B.~V.~Jacak and A.~Drees,
Eur.\ Phys.\ J.\ C {\bf 18} (2001) 645
[arXiv:nucl-th/9909075].

\bibitem{willis99}
Willis~N et al. [NA50 collaboration], Nucl. Phys.~{\bf A661}, (1999) 534c

%
\end{thebibliography}
\end{document}